%% file: Eobs_paper.tex
\documentclass[5p,longtitle]{elsarticle}  
\usepackage{graphicx}  
\usepackage{dcolumn}   
\usepackage{bm}        
\usepackage{amssymb}   

\usepackage{textcomp}
\usepackage{verbatim}
\usepackage{amsmath}
\usepackage{appendix}

\usepackage{enumitem}
\setdescription{leftmargin=4\parindent,labelindent=2\parindent}

\usepackage{hyperref}

\usepackage{epstopdf}
\providecommand*{\unit}[1]{\ensuremath{\mathrm{\,#1}}}

\def\ie{{\it i.e.~}}

\def\1/2{\frac{1}{2}}
\def\3/2{\frac{3}{2}}

\begin{document}

\begin{frontmatter} 

\title{First measurement of the helicity asymmetry $\mathbf E$ in $\boldsymbol \eta$ photoproduction on the proton }

\input{authors_final_els.tex}

\begin{abstract}
	Results are presented for the first measurement of the
        double-polarization helicity asymmetry $E$ for 
	the $\eta$ photoproduction reaction $\gamma p \rightarrow \eta p$. 
	Data were obtained using the FROzen Spin Target (FROST) with the CLAS spectrometer
        in Hall B at Jefferson Lab, covering a range of center-of-mass energy 
        $W$ from threshold to 2.15\unit{GeV} 
        and a large range in center-of-mass polar angle. 
        As an initial application of these data, the results have been 
        incorporated into the J\"ulich-Bonn model to examine the case 
        for the existence of a narrow $N^*$ resonance between 1.66 and 1.70\unit{GeV}.
        The addition of these data to the world database results in marked changes 
        in the predictions for the $E$ observable from that model. 
	Further comparison with several theoretical approaches indicates these data will 
        significantly enhance our understanding of nucleon resonances.	
\end{abstract}



\end{frontmatter}

\section{Introduction}

	Much activity is being devoted to establishing the details of 
        the excitation spectrum of the nucleon
	in order to deepen our understanding of that fundamental 
        strongly-interacting three-quark system. 
        Due to the broad widths of the nucleon excitations 
        (of the order of 100-300\unit{MeV}), 
        the states overlap in the mass spectrum. Thus, disentangling the 
        individual states to identify their exact masses and quantum numbers 
        has been quite difficult.        
	While some resonances are well established, fewer states have been observed
	than most constituent quark models and Lattice QCD calculations predict \cite{Klempt:2009pi}. 
	An additional complexity arises because, beyond resonance states with typical widths, 
	approaches based on chiral quark solitons also predict states with far
	narrower widths than do constituent quark models, including, 
        for example, an $N{\frac{1}{2}}^+$ 
        state with a width of 40\unit{MeV} or 
        less~\cite{Diakonov:1997mm,Diakonov:2003jj,Arndt:2003ga,Ellis:2004uz,Praszalowicz:2004dn} 
        at about 1.7\unit{GeV}; this particular state may have been 
        observed in $\eta$ photoproduction on the neutron~\cite{Anisovich:2013sva,Werthmuller:2013rba,PhysRevC.91.042201}.
	
         Since differential cross section measurements alone are insufficient to locate the 
        underlying resonance states or determine their properties, 
        attention has turned to polarization observables.
	Polarization observables involve interferences between sets of amplitudes,
	so their measurement can provide stringent tests for predictions of the
	photoproduction process and help sort out ambiguities in the theoretical
	description of the reaction in terms the resonances involved. 
	One such polarization observable is the helicity asymmetry $E$ 
        in pseudoscalar meson photoproduction,  
        which is the normalized difference in photoproduction yield when spins 
        of the incident photon and a longitudinally-polarized target 
        are parallel and anti-parallel.
	Formally, this observable is defined as a modulation 
        of the center-of-mass differential 
        cross section $d\sigma/d\Omega_0$ through the relation	
	\begin{equation}
		\frac{d\sigma}{d\Omega} = 
			\frac{d\sigma}{d\Omega}_0 \left(1 - P^T_z P^\gamma_\circ E \right),
		\label{eq:Edef}
	\end{equation}
	where $P^T_z$ specifies the degree of longitudinal target polarization and
	$P^\gamma_\circ$ is the circular polarization fraction of the incident photon beam.	
	This asymmetry is generally expressed as a function of the 
	center-of-mass energy $W$ and the polar angle of the produced meson
	in the center-of-mass frame $\cos\theta_{cm}$.

Pion photoproduction studies have contributed greatly 
to the knowledge of the nucleon resonance spectrum. 
Recent measurements of spin observables in pion photoproduction
\cite{Strauch:2005cs,Gottschall:2013uha,Krambrich:2009te,Wilson:2010zz,Oberle:2014rap},
have illustrated the power of polarization observables to clarify that spectrum. 
Even so, many ambiguities still exist and many predicted states remain unobserved. 
Though pion photoproduction offers a larger cross section, 
        the photoproduction of $\eta$ mesons 
	exhibits the interesting feature that the process excludes contributions 
        from resonances with isospin $I\!=\!3/2$,
        thereby isolating the $N^*(I\!=\!1/2)$ states. 
        The $\eta$ photoproduction process on the proton thus acts
	as an ``isospin filter'' for the nucleon resonance spectrum, 
        resulting in a useful tool for
	disentangling the different states,
and is especially important in finding and investigating states that do not couple strongly to pions.
	
\section{The Experiment}

	The measurements reported here
        are an integral part of a program at Jefferson Lab 
        to achieve a ``complete'' experiment for
        the $\eta$ photoproduction process, 
	whereby all the helicity amplitudes are determined
        for photoproduction of that pseudoscalar meson.
	The program began with measurements of the unpolarized 
        differential cross section $\frac{d\sigma}{d\Omega}_0$
	\cite{Dugger:2002ft,Williams:2009yj} using the 
        large solid angle CEBAF Large Acceptance Spectrometer (CLAS)~\cite{Mecking:2003zu},
	and the bremsstrahlung photon tagger housed in Jefferson Lab Hall B~\cite{Sober:2000we}.
	For the measurements reported here, 
	circularly polarized photon beams were produced 
	by polarization transfer from the polarized electron beam of the CEBAF accelerator,
	which was incident on an amorphous radiator of the photon tagger.

	The target nucleons for the photoproduction process were free protons  
        in frozen butanol ($\mathrm{C_4 H_9 OH}$) beads
	inside a 50-\unit{mm}-long target cup.~\cite{Keith:2012ad}.
        The protons of the hydrogen atoms in this material were 
        dynamically polarized along the photon beam direction.
	The longitudinal target polarization $P^T_z$ was determined with
	nuclear magnetic resonance measurements, and
	averaged $82\pm5\%$.
 	To minimize systematic uncertainties, the orientation of the 
        target polarization direction was flipped every few days of data-taking between 
        being aligned and anti-aligned with the direction of the incoming photon beam.
	The helicity of the beam was flipped at a rate of 30\unit{Hz}. 
        
	Final state particles resulting from 
        photoproduction were detected using CLAS,
	a set of six identical charged particle detectors 
	installed in a toroidal magnetic field.
	The principal CLAS subsystems required for this study were 
        the drift chamber systems for 
	tracking charged particles~\cite{Mestayer:2000we},
	a scintillator-based time-of-flight system~\cite{Smith:1999ii}, and
	a start counter array which determined when charged particles passed from
	the target into the detection region~\cite{Sharabian:2005kq}. 
	The energy and polarization information for incident photons
        was provided by the photon tagger.

\section{Analysis}


	To determine the helicity asymmetry $E$ in a discrete event counting 
	experiment, Eq.~(\ref{eq:Edef}) is inverted to form the asymmetry  
	\begin{equation}\displaystyle
		E = -\frac{1}{|P^T_z| |P^\gamma_\circ|}
			\left(\frac{N_+ - N_-}{N_+ + N_-}\right),
                \label{eq:EfromCounts}
	\end{equation}
	where the detector acceptance cancels. 
        The cross sections are replaced by $N_+$ and $N_-$, which are
        the number of $\eta$ mesons counted in beam-target helicity 
        aligned and anti-aligned settings, respectively. 
        The background from non-$\eta$ final states and those from events arising 
	from the unpolarized nucleons within the butanol are subtracted
        before forming this asymmetry.

	Determination of the $E$ observable requires knowledge 
        of the degree of polarization for both the photon beam and the target proton.  
	The photon beam polarization is calculated from the incident photon energy $E_\gamma$
	relative to the bremsstrahlung endpoint ($\tilde{E}=E_\gamma/E_{e^-}$)
	via the expression 
	\begin{equation}
		P^\gamma_\circ = P_e\frac{4\tilde{E} - \tilde{E}^2}{4-4\tilde{E}+3\tilde{E}^2},
		\label{eq:circPol}
	\end{equation}
	where $P_e$ is the polarization of the 
	electron beam incident on the amorphous radiator within the photon tagger~\cite{Olsen:1959zz}; 
	$P_e$ was measured with the Hall B M{\o}ller polarimeter 
        during the experiment to be $0.84\pm0.01$.


	Events in the detector were reconstructed in the following manner. 
        Individual charged tracks were reconstructed in the CLAS drift chambers 
        and matched to hits in the time-of-flight (TOF) and start counter paddles.
	The particle identity was determined by combining the information on the 
	momentum of the particle, which was
        determined by the drift chambers from the curvature of the 
        particle trajectory in the magnetic field, 
        and on the speed of the particle ($\beta$) as determined from the timing information
        provided by the tagger, start counter, and TOF systems.
	Charged tracks that could not be reconstructed by 
	all of these detectors were rejected.
	A track was assumed to have the
	particle identity that allowed the closest match between the 
        4-momentum-computed $\beta$ and the measured value of $\beta$.
	An additional requirement that the measured $\beta$ was within 
        $\pm0.04$ of the expected 
	value was imposed on pion candidates, 
	significantly suppressing the electron background.
	Once the particle identity was established, a
	correction due to energy loss in the target and detector materials was performed,
	with the 4-vector values adjusted accordingly.
        The tracks and the event as a whole were associated to beam photons
        based on consistency with the projected vertex timing.
	To avoid ambiguity, only events with particles matching 
        exactly one beam photon were kept.


        The CLAS detector is primarily a 
	charged particle spectrometer, with electromagnetic calorimetry 
	confined to a narrow angular range. Thus, $\sim94\%$ of the signal in
	this analysis relied on missing mass reconstruction of the neutral $\eta$
        from the measured kinematical information of the proton 
	recoiling into the CLAS (the detection of which was required),
	with the remainder of events having one or both charged pions
        from the decay  $\eta \rightarrow \pi^+\pi^-\pi^0$ detected.
        Events with a single detected charged pion were required
	to have a missing mass squared greater than $0.06\unit{GeV^2/c^4}$,
	which is the onset of the remaining two-pion phase space.
	Events with both a $\pi^+$ and $\pi^-$ detected were required to have 
        the remaining missing mass squared close to that of the $\pi^0$ 
	within the detector resolution: $0.008 - 0.028\unit{GeV^2/c^4}$.


	The $\eta$ photoproduction data were analyzed to extract the
	helicity asymmetry $E$ in 50\unit{MeV}-wide 
	center-of-mass energy $W$ bins and 0.2-wide 
	center-of-mass production $\eta$ polar angle ($\cos\theta_\mathrm{cm}$) bins.
	Binning in $W$ begins near the $\eta$ threshold at $1.5\unit{GeV}$.
	These bin widths were chosen to balance between minimizing statistical uncertainties for the
	extraction while achieving the best energy resolution for the 
	resonance spectrum and most thorough knowledge of the polar distribution of the resonance decay.
	The analysis procedures described below were performed for each kinematic bin
	separately.


	\begin{figure*}[!htb]\centering
          \includegraphics[width=2.2in]{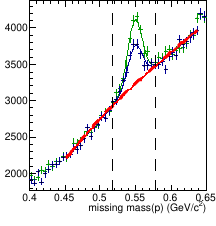}\hspace{0.1in}
          \includegraphics[width=2.2in]{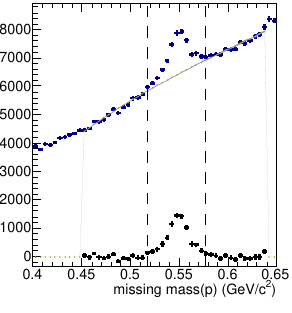}\hspace{0.1in}
          \includegraphics[width=2.2in]{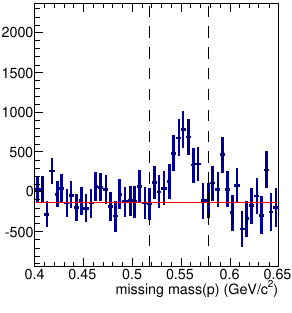} 
	  \caption{Analysis example for the kinematic bin 
            ($1650<W<1700\unit{MeV}, -0.2<\cos\theta_\mathrm{cm}<0.0$).
	    Left panel: Background fit to the missing mass spectra for the 
            two helicities (higher amplitude $N_-$ is green). 
          Middle panel: Background subtraction and net $\eta$ 
	    yield.
          Right panel: Yield difference, 
		with fit to sidebands to determine overall asymmetry offset. (Color online)
	    \label{EanaBinEx}}
	\end{figure*}

	To distinguish a photoproduced $\eta$ from the background, 
	fits were performed to the invariant mass spectra with models of the signal
	and background included, as shown in Fig.~\ref{EanaBinEx}. 
        The integral of the fit shape of the background 
	and the uncertainty of this integral from the error matrix estimated 
        the background contribution. Since the detector resolution 
        dominated the shape of the $\eta$ enhancement,
	the signal was modeled as a Gaussian. 
        Polynomials were used to model the background,
	with the order of the polynomial increased progressively 
        with every fit iteration up to fifth order as long as the 
        fit improvement was statistically significant. 	
	Specifically, improvement in the confidence level beyond 0.5 
	was considered not significant.
	For the two $W$ bins near the $\eta$ threshold ($W<1.6\unit{GeV}$),
	the step function-like drop-off in photoproduced system phase space required 
	a different approach. For those bins, the error function $\mathrm{erf}$ was used in addition
	to the polynomial, with its amplitude and transition width as free parameters.

	A single fit was performed to the spectra of both beam helicities, 
        with a common model of the background shape and common position and width
        of the $\eta$ enhancement; an example is shown in the left panel of
        Fig.~\ref{EanaBinEx}.
	The unpolarized background essentially cancels out 
in the difference of the yield in the missing mass spectra of the two 
        helicities ($N_+$ and $N_-$) seen in the numerator of Eq.~(\ref{eq:EfromCounts});
an example is given in the middle panel of Fig.~\ref{EanaBinEx}.
	The small remaining overall vertical offset seen
        for some kinematic bins may be due to asymmetries
	in the broad polarized background, 
        such as the asymmetry which might exist for the $\pi^+\pi^-$
	final state. These offsets were determined with a fit to the 
	sidebands (as seen in the right panel of Fig.~\ref{EanaBinEx}),
	defined as $\pm 3\sigma$ from the peak center, where the center
	and $\sigma$ values were derived from the previously performed fit
	of the  $\eta$ enhancement. The normalized asymmetry was thus calculated
	from this corrected difference of helicities divided by the
	overall $\eta$ yield determined with the background subtraction
	described above. 


        A separate study of photoproduction on a 
        pure carbon target showed no evidence of peaking in the $\eta$ mass region.
        Thus, no correction for the heavier nuclei in the target was required
        beyond the smooth background fitting described.
        Helicity asymmetry extraction was not performed
	when the background exhibited an extremum under the $\eta$ peak, 
	(\ie within $\pm1\sigma$ of the  peak centroid) to avoid serious 
        ambiguities between the signal and the background shape.
        Additionally, analysis in a kinematic bin was abandoned when
        the total $\eta$ yield uncertainty was greater than 30\%.

        Missing mass energy resolution for the $\eta$ with CLAS 
        is a smooth function of the kinematic 
        space explored here. Therefore, the peak widths seen in the 
        initial independent analyses 
        of the individual kinematic bins were compared and 
        a smooth function of the peak width across the kinematic space
        was extracted. Yield extractions were then repeated using these constraints 
        on the peak width to enforce consistency with the detector resolution.

	Statistical uncertainties dominated the systematic uncertainties in all analyzed bins,
      and are shown combined in the presented results.
	The systematic uncertainties include the target polarization $P^T_z$ uncertainty 
	6.1\%) and photon beam polarization $P^\gamma_\circ$ uncertainty (3.1\%). 

\section{Results and discussion}

\begin{figure}[!htb]\hspace{-0.08in}
	  \includegraphics[width=3.75in]{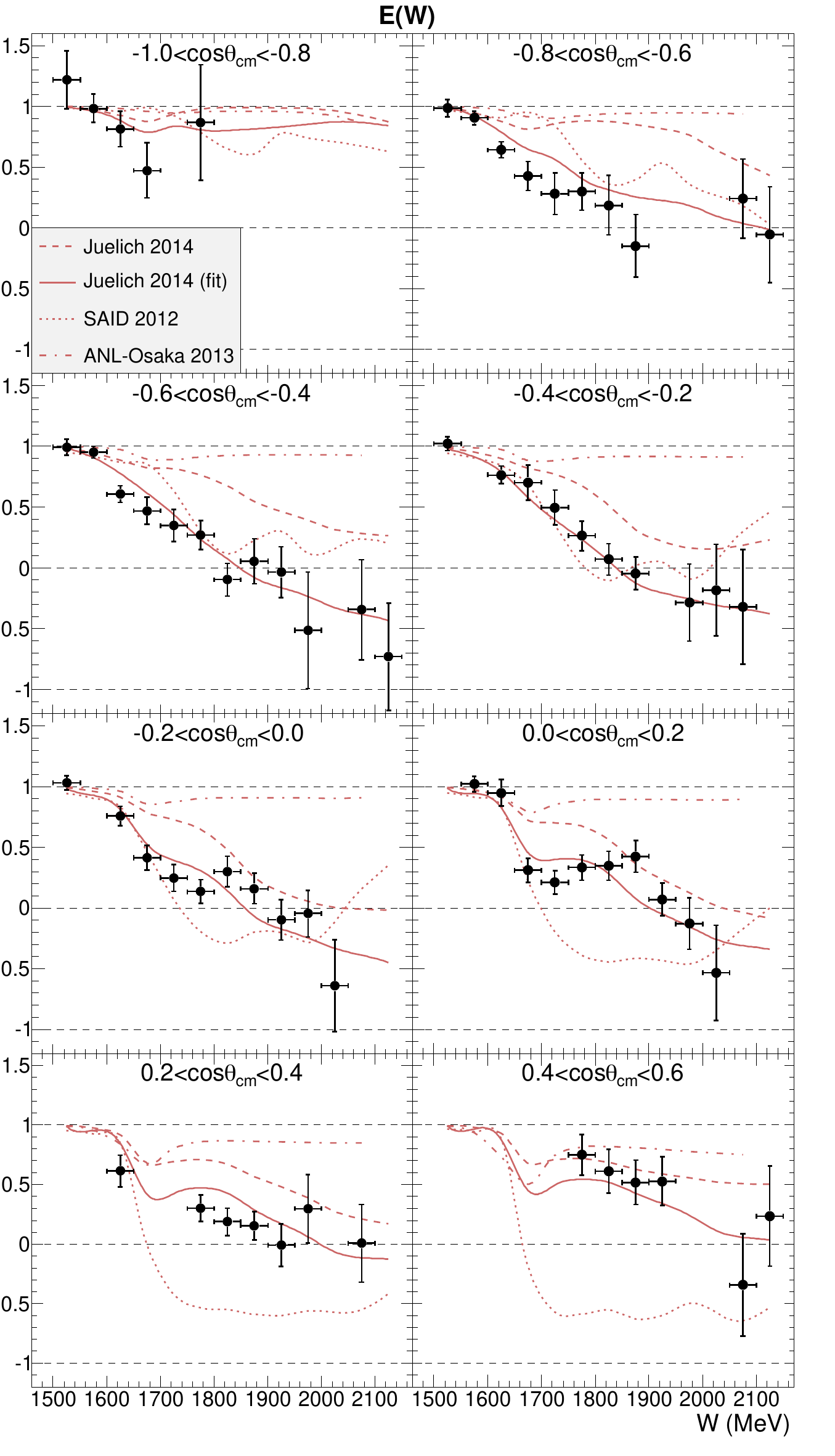}
	  \caption{Helicity asymmetry $E$ for
	    $\gamma p \rightarrow \eta p$ as a function of
	    $W$ at various values of $\cos\theta_\mathrm{CM}$ compared
           to several phenomenological predictions.
	    \label{Eresults2}
          }          
	\end{figure}

\begin{figure}[!htb]\hspace{-0.12in}
	  \includegraphics[width=3.8in]{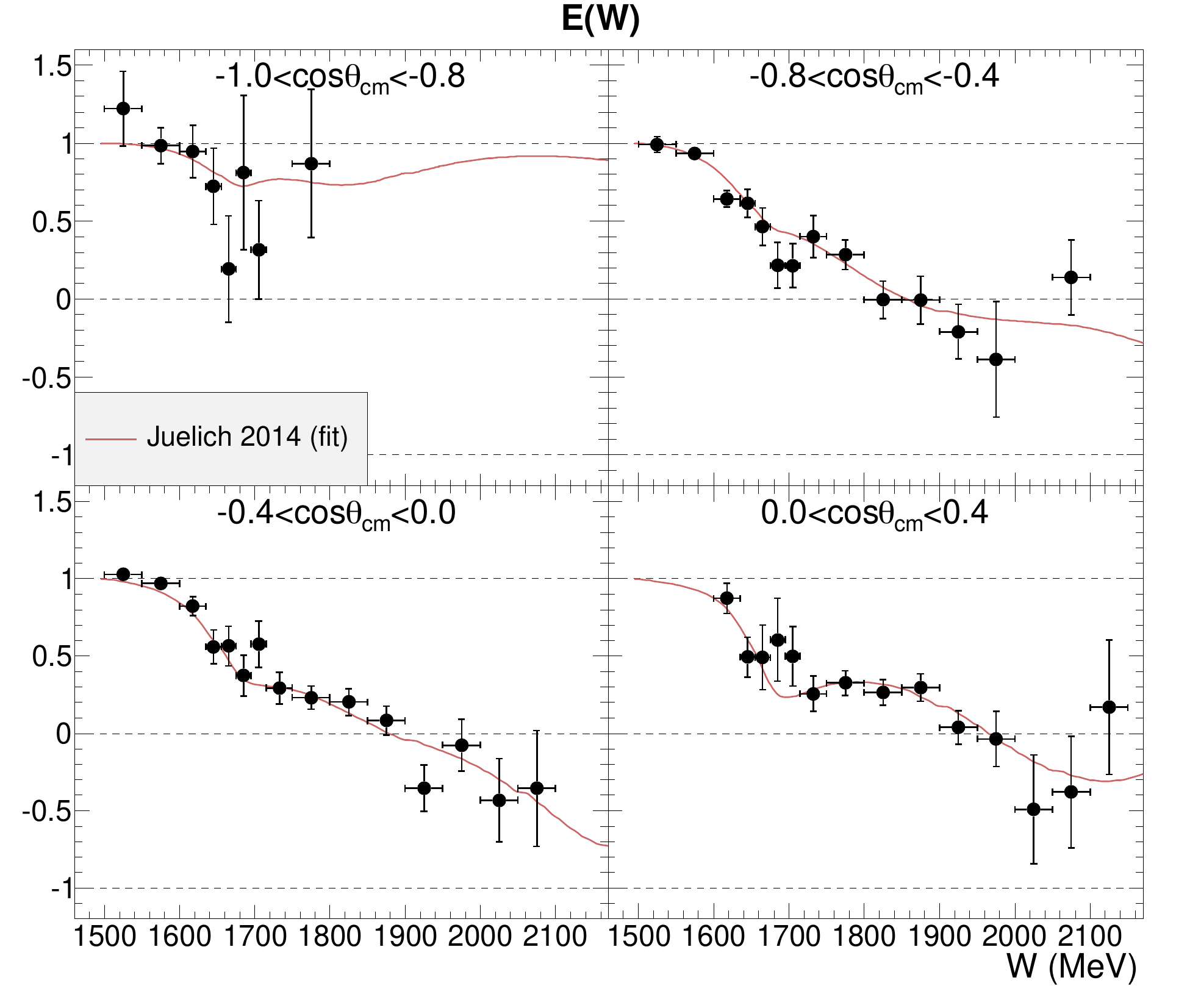}
	  \caption{The helicity asymmetry $E$ for 
	    $\gamma p \rightarrow \eta p$ using smaller $W$
	    bins 
           to explore the behavior of the helicity observable near $W\sim1.7$ GeV.
	Predictions of the J\"ulich-Bonn model as discussed in the text are shown by the solid line. 
	    \label{Eresults3}
          }
	\end{figure}

	The results for the helicity asymmetry $E$ are
	shown in Figs.~\ref{Eresults2} and \ref{Eresults3} for 1.5 $\le W \le$ 2.1\unit{GeV}. 
        At threshold, the $E$ observable is close to unity due to the dominance of the 
        $N(1535)\frac{1}{2}^-$ resonance~\cite{Hejny:1999iw},  
	and the results reported here are consistent 
	within uncertainties with this expectation. 
	As $W$ increases, the presence of other resonances and the interferences of
	the various amplitudes related to those resonances generate a $W$-dependent
	structure in $E$, which models of the production process attempt to describe.
	As examples of such models, 
	shown in Fig.~\ref{Eresults2} are predictions from phenomenological fits 
	by 
        SAID~\cite{McNicoll:2010qk}, 
	J\"ulich-Bonn~\cite{Ronchen:2015vfa} 
        and ANL-Osaka~\cite{Kamano:2013iva}. 
        
        The figure also shows a new fit with the J\"ulich-Bonn dynamical 
        coupled-channel approach incorporating the data reported here. 
        In that framework, 
        the hadronic scattering amplitude is constructed with a potential, generated 
        from an effective SU(3) Lagrangian, using time-ordered perturbation theory, and 
        the amplitude is iterated in a Lippmann-Schwinger equation such that 
        unitarity and analyticity are automatically respected. 

        This new fit also simultaneously incorporated the world databases for
        the pion-induced production of $\eta N$, $K\Lambda$ and $K\Sigma$ final
        states~\cite{Ronchen:2012eg} and the partial-wave solution of the GW/DAC
        group~\cite{Workman:2012hx} for elastic $\pi N$ scattering. It also
        includes the world data bases of pion and $\eta$ photoproduction off the
        proton up to $W~\sim 2.3$~GeV~\cite{Ronchen:2014cna, Ronchen:2015vfa}, in
        particular the recent MAMI results on $T$ and $F$ in $\eta$
        photoproduction~\cite{Akondi:2014ttg}. 
	
        This new fit also simultaneously incorporated measurements for 
        pion-induced reactions and for pion- and $\eta$ photoproduction off the proton. 
        The published J\"ulich-Bonn predictions for $E$, as well as this new fit, 
        include the recent MAMI results on 
        $T$ and $F$ in $\eta$ photoproduction~
        
        In order to achieve a good fit result, all parameters tied 
        to the resonance states and to the photon interaction had to be modified from 
        the values reported in Ref.~\cite{Ronchen:2015vfa}.
	The inclusion of these new $E$ data also resulted in significant changes 
        in the extracted resonance pole positions.
        For example, with these new $E$ data, the $N(1710)1/2^+$ resonance
        becomes 45 \unit{MeV} heavier and 20 \unit{MeV} wider compared to 
        Ref.~\cite{Ronchen:2015vfa}, with a 40\% smaller branching ratio into the
        $\eta N$ channel. Helicity couplings for the high-spin
        $N(2190)7/2^-$ and $N(2250)9/2^-$ resonances, whose properties are
        difficult to determine in general, change their pole positions by up to
        80\unit{MeV} in the real and 100\unit{MeV} in their imaginary parts due to
        these new $E$ data.

        That the E data reported here have such a large impact on the resonance 
        parameters might appear surprising. Since the different spin observables 
        have differing combinations of amplitudes, the various observables will 
        have differing degrees of impact on reducing the uncertainties of the 
        parameter values extracted from a fit to the experimental data. 
        In the present case, those parameters are the fundamental 
        electromagnetic properties of resonances, 
        their helicity couplings at the pole. 

        To study the variations in the statistical impacts on the parameters of 
        the new J{\"u}lich fit that arise through the use of measurements of the 
        different spin observables, we have studied the condition number 
        $\kappa$ for the covariance matrix found in fitting the observed data. 
        The condition number $\kappa$ is a standard test for diagnosing 
        multicollinearity and, hence, the non-orthogonality of the model 
        parameters on which the fit is based \cite{kappa}. 
        The condition number $\kappa$ is defined as 
        $\lambda_{\rm max}/\lambda_{\rm min}$, that is, the ratio of the 
        largest and smallest eigenvalues. Geometrically, $\sqrt{\kappa}$ 
        determines the ratio of the longest half-axis of the statistical 
        uncertainty ellipsoid divided by the shortest one. 
        A large $\kappa$ (say, greater than 100 \cite{kappa}) 
        is a sign of moderate to strong multicollinearity, 
        \ie a very elongated statistical uncertainty ellipsoid; 
        larger values of $\kappa$ thus connote greater uncertainty in the 
        corresponding helicity couplings determined from the fit. 

        For the present $E$ data, we found $\kappa\approx 50$, 
        while for the other spin observables 
        (and even the differential cross section), $\kappa$ ranged from 
        $50$ to $400$. Thus, in terms of minimizing the uncertainties 
        in the extracted parameters, the $E$ observable measurements 
        reported here indeed turn out to be particularly impactful. 
        This underscores that the observable $E$ in $\eta$ photoproduction
        is especially suited to disentangle electromagnetic resonance 
        properties. With relatively few data points, 
        this measurement offers a larger impact on the baryon spectrum, 
        helicity couplings, and even hadronic decay parameters 
        than might be expected.        

 
%

	Turning next to the putative $N{\frac{1}{2}}^+$ resonance near $W\sim 1.7$ GeV, 
        Fig.~\ref{Eresults3} shows our results for the observable $E$ 
	using finer
	$W$ bins (of 20\unit{MeV} width). 
	Coarser, 0.4-wide binning in $\cos\theta_\mathrm{cm}$ was used to compensate 
        for the narrow energy binning. 
        A fit to this re-binned 
        data using the J\"ulich-Bonn formalism found that the structure observed at $\sim1.7\unit{GeV}$
        for the $\cos\theta_\mathrm{cm}$ bin centered at $0.2$ 
        is due to interference between the $E_0^+$ and $M_1^+$ multipoles, which
        vary rapidly at this energy due to the $N(1650)1/2^-$ and 
        $N(1720)3/2^+$ resonances. Together with the slowly varying
        $E_2^-$ multipole, these three multipoles alone describe the $E$ asymmetry quite well
	without the need for an additional narrow resonance near 1.68\unit{GeV}. 
        A similar analysis of the multipole content for the 
	$\cos\theta_\mathrm{cm}$ bin centered at $-0.6$ shows that the interference of the 
        $E_0^+$ and $M_2^+$ multipoles (the latter containing the $N(1675)5/2^-$) 
        is responsible for the dip, with $E_1^+$, $E_2^-$ and $M_2^-$ necessary to
        better approximate the full fit.  
	Combined with the hints seen in Refs.~\cite{Anisovich:2013sva,Werthmuller:2013rba}, 
        the data presented here further 
	motivate additional experimental investigations looking at other spin observables.

	In summary, we have presented the first measurements of the helicity asymmetry $E$
	in photoproduction of $\eta$ mesons from the proton.
	Initial investigation of these results with the J\"ulich-Bonn dynamical coupled-channel approach
	show pronounced changes in the description of this variable when these new
	data are included, and demonstrate how
	these measurements are particularly impactful in constraining analyses of the excitation 
        spectrum of the proton. With respect to the existence of an        
        $N\frac{1}{2}^+$ resonance near $W\sim1.7$ GeV
        suggested previously~\cite{Diakonov:1997mm,Diakonov:2003jj,Arndt:2003ga,Ellis:2004uz,Praszalowicz:2004dn},
	the data obtained here do not demand the presence of such a state, 
	but further measurements of other polarization observables would be helpful in gaining
	additional insight on that question.

\section*{Acknowledgements}	
	The authors gratefully acknowledge the work of the Jefferson Lab staff,
        as well as the support by the National Science Foundation,
        the JSC(JUROPA) facility at FZ J\"ulich,
        the French Centre National de la Recherche Scientifique and 
        Commissariat \`a l'Energie Atomique, 
        the Italian Istituto Nazionale di Fisica Nucleare, 
        the Chilean Comisi\'on Nacional de Investigaci\'on Cient\'ifica y Tecnol\'ogica,
        the Science and Technology Facilities Council of the United Kingdom,
        and the National Research Foundation of Korea.
        This material is based upon work supported by the 
        U.S. Department of Energy, Office of Science, 
        Office of Nuclear Physics under contract DE-AC05-06OR23177.

\section*{References}
        \bibliography{ref}

\end{document}

%% file: authors_final_els.tex
\newcommand*{\ANL}{Argonne National Laboratory, Argonne, Illinois 60439}
\newcommand*{\ANLindex}{1}
\newcommand*{\ASU}{Arizona State University, Tempe, Arizona 85287-1504}
\newcommand*{\ASUindex}{2}
\newcommand*{\CSUDH}{California State University, Dominguez Hills, Carson, CA 90747}
\newcommand*{\CSUDHindex}{3}
\newcommand*{\CANISIUS}{Canisius College, Buffalo, NY}
\newcommand*{\CANISIUSindex}{4}
\newcommand*{\CMU}{Carnegie Mellon University, Pittsburgh, Pennsylvania 15213}
\newcommand*{\CMUindex}{5}
\newcommand*{\CUA}{Catholic University of America, Washington, D.C. 20064}
\newcommand*{\CUAindex}{6}
\newcommand*{\SACLAY}{CEA, Centre de Saclay, Irfu/Service de Physique Nucl\'eaire, 91191 Gif-sur-Yvette, France}
\newcommand*{\SACLAYindex}{7}
\newcommand*{\UCONN}{University of Connecticut, Storrs, Connecticut 06269}
\newcommand*{\UCONNindex}{8}
\newcommand*{\FU}{Fairfield University, Fairfield CT 06824}
\newcommand*{\FUindex}{9}
\newcommand*{\FIU}{Florida International University, Miami, Florida 33199}
\newcommand*{\FIUindex}{10}
\newcommand*{\FSU}{Florida State University, Tallahassee, Florida 32306}
\newcommand*{\FSUindex}{11}
\newcommand*{\GWUI}{The George Washington University, Washington, DC 20052}
\newcommand*{\GWUIindex}{12}
\newcommand*{\HISKP}{HISKP and BCTP, Universit\"at Bonn, 53115 Bonn, Germany}
\newcommand*{\HISKPindex}{12.5}
\newcommand*{\ISU}{Idaho State University, Pocatello, Idaho 83209}
\newcommand*{\ISUindex}{13}
\newcommand*{\INFNFE}{INFN, Sezione di Ferrara, 44100 Ferrara, Italy}
\newcommand*{\INFNFEindex}{14}
\newcommand*{\INFNFR}{INFN, Laboratori Nazionali di Frascati, 00044 Frascati, Italy}
\newcommand*{\INFNFRindex}{15}
\newcommand*{\INFNGE}{INFN, Sezione di Genova, 16146 Genova, Italy}
\newcommand*{\INFNGEindex}{16}
\newcommand*{\INFNRO}{INFN, Sezione di Roma Tor Vergata, 00133 Rome, Italy}
\newcommand*{\INFNROindex}{17}
\newcommand*{\INFNTUR}{INFN, Sezione di Torino, 10125 Torino, Italy}
\newcommand*{\INFNTURindex}{18}
\newcommand*{\ORSAY}{Institut de Physique Nucl\'eaire, CNRS/IN2P3 and Universit\'e Paris Sud, Orsay, France}
\newcommand*{\ORSAYindex}{19}
\newcommand*{\ITEP}{Institute of Theoretical and Experimental Physics, Moscow, 117259, Russia}
\newcommand*{\ITEPindex}{20}
\newcommand*{\JMU}{James Madison University, Harrisonburg, Virginia 22807}
\newcommand*{\JMUindex}{21}
\newcommand*{\JUELICH}{Forschungszentrum J\"ulich, Institut f\"ur Kernphysik, 52425 J\"ulich, Germany}
\newcommand*{\JUELICHindex}{21.5}
\newcommand*{\KNU}{Kyungpook National University, Daegu 702-701, Republic of Korea}
\newcommand*{\KNUindex}{22}
\newcommand*{\UNH}{University of New Hampshire, Durham, New Hampshire 03824-3568}
\newcommand*{\UNHindex}{23}
\newcommand*{\NSU}{Norfolk State University, Norfolk, Virginia 23504}
\newcommand*{\NSUindex}{24}
\newcommand*{\OHIOU}{Ohio University, Athens, Ohio 45701}
\newcommand*{\OHIOUindex}{25}
\newcommand*{\ODU}{Old Dominion University, Norfolk, Virginia 23529}
\newcommand*{\ODUindex}{26}
\newcommand*{\RPI}{Rensselaer Polytechnic Institute, Troy, New York 12180-3590}
\newcommand*{\RPIindex}{27}
\newcommand*{\URICH}{University of Richmond, Richmond, Virginia 23173}
\newcommand*{\URICHindex}{28}
\newcommand*{\ROMAII}{Universita' di Roma Tor Vergata, 00133 Rome Italy}
\newcommand*{\ROMAIIindex}{29}
\newcommand*{\MSU}{Skobeltsyn Institute of Nuclear Physics, Lomonosov Moscow State University, 119234 Moscow, Russia}
\newcommand*{\MSUindex}{30}
\newcommand*{\SCAROLINA}{University of South Carolina, Columbia, South Carolina 29208}
\newcommand*{\SCAROLINAindex}{31}
\newcommand*{\TEMPLE}{Temple University, Philadelphia, PA 19122 }
\newcommand*{\TEMPLEindex}{32}
\newcommand*{\JLAB}{Thomas Jefferson National Accelerator Facility, Newport News, Virginia 23606}
\newcommand*{\JLABindex}{33}
\newcommand*{\UTFSM}{Universidad T\'{e}cnica Federico Santa Mar\'{i}a, Casilla 110-V Valpara\'{i}so, Chile}
\newcommand*{\UTFSMindex}{34}
\newcommand*{\EDINBURGH}{Edinburgh University, Edinburgh EH9 3JZ, United Kingdom}
\newcommand*{\EDINBURGHindex}{35}
\newcommand*{\GLASGOW}{University of Glasgow, Glasgow G12 8QQ, United Kingdom}
\newcommand*{\GLASGOWindex}{36}
\newcommand*{\VT}{Virginia Tech, Blacksburg, Virginia 24061-0435}
\newcommand*{\VTindex}{37}
\newcommand*{\VIRGINIA}{University of Virginia, Charlottesville, Virginia 22901}
\newcommand*{\VIRGINIAindex}{38}
\newcommand*{\WM}{College of William and Mary, Williamsburg, Virginia 23187-8795}
\newcommand*{\WMindex}{39}
\newcommand*{\YEREVAN}{Yerevan Physics Institute, 375036 Yerevan, Armenia}
\newcommand*{\YEREVANindex}{40}
\newcommand*{\WATERLOO}{University of Waterloo, Institute for Quantum Computing, Waterloo, Ontario, Canada}
\newcommand*{\WATERLOOindex}{41}

\newcommand*{\NOWUK}{University of Kentucky, Lexington, Kentucky 40506}
\newcommand*{\NOWJLAB}{Thomas Jefferson National Accelerator Facility, Newport News, Virginia 23606}
\newcommand*{\NOWODU}{Old Dominion University, Norfolk, Virginia 23529}
\newcommand*{\NOWEDINBURGH}{Edinburgh University, Edinburgh EH9 3JZ, United Kingdom}

\author[toASU]{I.~Senderovich}\ead{senderov@jlab.org}
\author[toASU]{B.T.~Morrison}
\author[toASU]{M.~Dugger}
\author[toASU]{B.G.~Ritchie}
\author[toASU]{E.~Pasyuk} 
\author[toASU]{R.~Tucker}

\author[toJLAB]{J.~Brock}
\author[toJLAB]{C.~Carlin}
\author[toJLAB]{C.D.~Keith}
\author[toJLAB]{D.G.~Meekins}
 
% ...
\address[toASU]{\ASU} 
% ...
\address[toJLAB]{\JLAB} 
% ...
 
 \fntext[toNOWJLAB]{Current address: Newport News, Virginia 23606 }
 \fntext[toNOWUK]{Current address: LEXINGTON, KENTUCKY 40506 }
 \fntext[toNOWODU]{Current address: Norfolk, Virginia 23529 }
 \fntext[toNOWEDINBURGH]{Current address: Edinburgh EH9 3JZ, United Kingdom }